\def\cpt07{CPT~'07}
\def\al{\alpha}
\def\be{\beta}
\def\ga{\gamma}

\def\ep{\epsilon}

\def\ka{\kappa}
\def\la{\lambda}

\def\si{\sigma}
\def\ta{\tau}

\def\ps{\psi}

\def\Om{\Omega}

\def\cL{{\cal L}}
\def\etall{{\it et al.}}
\def\kt{{\tilde\ka}}
\def\bt{{\tilde b}}
\def\sb{\overline s}
\newcommand{\beq}{\begin{equation}}
\newcommand{\eeq}{\end{equation}}
\newcommand{\bea}{\begin{eqnarray}}
\newcommand{\eea}{\end{eqnarray}}

\documentclass[12pt]{iopart}
\begin{document}

\title{Atomic and optical tests of Lorentz symmetry
}

\author{Neil Russell}

\address{Physics Department,
Northern Michigan University,
Marquette, MI, USA}
\ead{nrussell@nmu.edu}
\begin{abstract}
This article\footnote{Invited Comment for CAMOP section of {\it Physica Scripta}}
 reports on the
Fourth Meeting on Lorentz and CPT Symmetry,
CPT '07,
held in August 2007 in Bloomington, Indiana, USA.
The focus is on recent tests of Lorentz symmetry
using atomic and optical physics.
Results presented at the meeting
include
improved bounds on Lorentz violation in the photon sector,
and the first bounds on several coefficients in the gravity sector.
\end{abstract}

\section{Introduction}
The AMO community has played a major role
in testing Lorentz symmetry over the last decade.
Much of this is due
the innovative design work
of experimentalists,
who have steadily improved the attainable
levels of precision in various experiments.
Exquisite tests of Lorentz symmetry
in AMO and other areas
have been performed with
optical and microwave cavities
\cite{cavity,hm},
with atomic clocks and masers
\cite{clockmaser},
with torsion pendula
\cite{heckel06,tpend},
and with Penning traps confining electrons or protons
\cite{penntrap}.
These endeavors have vigorously sought
to test whether nature is exactly Lorentz symmetric.

The creation in the 1990s
of a broad theoretical framework
for Lorentz violation
 is a major reason for the surge of interest
in studies of Lorentz symmetry.
This framework,
which spans the spectrum of quantum and gravitational physics,
is called the Standard-Model Extension, or SME
\cite{dcak97}.
It has opened numerous avenues
to probe Planck-scale physics,
where Lorentz violations may occur,
without the need to attain the
$10^{19}$~GeV energies at which
the theories of particle physics and gravitation are expected to merge.

Lorentz violations are possible, for example,
in string theory with spontaneous symmetry
breaking.
This idea,
of using a potential to spontaneously break Lorentz symmetry,
thus enforcing a nonzero vacuum value for a tensor field,
was introduced by Kosteleck\'y and Samuel \cite{theory}.
Several models for such fields have been created as useful test cases,
and include the ones known as
the bumblebee field and the cardinal field \cite{modes}.
It is remarkable that
AMO experiments,
such as the ones mentioned above,
are able in principle to achieve sensitivity
to the vacuum expectation values of these fields.
Alternative approaches to Lorentz violation include ones involving
spacetime-varying fields \cite{spacetimevarying},
noncommutative field theories \cite{ncqed},
quantum-gravity \cite{qg},
branes \cite{brane},
supersymmetry \cite{mbak},
and a variety of others \cite{models}.

The goal of this article
is to provide background and details of
experimental and theoretical work
on Lorentz violation
in the area of AMO physics
presented at the Fourth Meeting on CPT and Lorentz Violation,
held in Bloomington, Indiana, in August 2007
\cite{cpt07}.
A number of new bounds on SME coefficients
were presented at the meeting.
A full listing of the experimental measurements of the SME coefficients
in all sectors can be found in Ref.\ \cite{LVtables}.

\section{The Standard-Model Extension}
The SME is defined at the level of the
effective field-theory action $S_{\mbox{\footnotesize SME}}$,
with a variety of Lorentz-violating terms appearing in the lagrangian density
$\cL_{\mbox{\footnotesize SME}}$:
\beq
S_{\mbox{\footnotesize SME}} = \int \cL_{\mbox{\footnotesize SME}} \, d^4x \,  .
\eeq
One way to evaluate the content of the lagrangian density $\cL_{\mbox{\footnotesize SME}}$
is to separate
the gravity and matter sectors:
\beq
\cL_{\mbox{\footnotesize SME}} = \cL_{\mbox{\footnotesize matter}} + \cL_{\mbox{\footnotesize gravity}}
\, .
\eeq
Terms in the gravitational piece
$\cL_{\mbox{\footnotesize gravity}}$
are constructed only from the basic gravitational fields.
The choice of these is guided by the need for a
realistic description of nature,
which must include particles with spin.
Since
Riemann-Cartan spacetimes incorporate spinors
in curved spacetimes
and can be constructed using
the vierbein
and the spin connection \cite{akgrav04},
these are the chosen basic fields.
The familiar gravitational fields,
such as the curvature and the torsion,
can be expressed in terms of
the vierbein and the spin connection.
The matter piece
$\cL_{\mbox{\footnotesize matter}}$
consists of all other terms.
It includes ones constructed from
the spinors $\ps$ describing ordinary matter (protons, neutrons, and electrons),
gauge fields such as $A^\mu$ for the photon,
and
the fields describing particles
that are not `ordinary,'
like muons, mesons, neutrinos and so on.
The terms in
$\cL_{\mbox{\footnotesize matter}}$
can include the basic gravitational fields together with these matter fields,
whereas
$\cL_{\mbox{\footnotesize gravity}}$
is `pure,'
containing only gravitational fields.

Lorentz violation in the flat-spacetime (Minkowski) limit
with no torsion
has been studied extensively
since the basic theoretical  framework was introduced
\cite{dcak97}.
In this limit, the metric $g_{\mu\nu}$
has nonzero constant values on the diagonal only,
there are no gravitational fields to consider,
and the only lagrangian density of relevance
is  $\cL_{\mbox{\footnotesize matter}}.$
The Lorentz-preserving part of this lagrangian density is
the standard model of particle physics,
while the Lorentz-violating part
contains terms with coefficients that
can be experimentally probed.

For more than a decade,
experimental limits have been placed on
coefficients for Lorentz violation in the
torsion-free Minkowski limit of the SME.
In the case of ordinary matter
and radiation,
relevant studies include
the ones mentioned above,
as well as others involving
high-speed ions
\cite{Reinhardt07},
cosmological birefringence
\cite{akmm01,akmm07},
and satellite-mounted oscillators \cite{bklr2003}.
For other particles and fields in the Minkowski limit of the matter sector,
theoretical and experimental studies
include ones looking at
muons \cite{bennett07,muons},
neutral mesons \cite{mesons},
neutrinos \cite{neutrino},
the Higgs \cite{higgs},
and baryogenesis \cite{baryogen}.
A variety of astrophysical processes
involving both ordinary and other matter
place limits on Lorentz violation \cite{ba}.

In the case of nonzero torsion,
the Minkowski limit of the matter sector
has recently been studied.
Several new bounds
on components of the torsion
tensor
have been found \cite{aknrjt}
based on experiments in the AMO field.

In the pure-gravity sector,
the Lorentz-preserving part of
$\cL_{\mbox{\footnotesize gravity}}$
contains the conventional Einstein-Hilbert lagrangian
from which the Einstein field equations follow
when torsion is zero and no other terms are present.
The Lorentz-violating terms in
$\cL_{\mbox{\footnotesize gravity}}$
provide a framework for a variety of experimental tests of
Lorentz symmetry in the context of pure gravity \cite{qbak06}.
The first experimental measurements of coefficients for
Lorentz-violating terms in this sector
were presented at \cpt07
\cite{battat,muller07}.

The following sections provide
an overview of recent experiments and theory
in each of these sectors,
with emphasis on results
relating to AMO physics that were
presented at \cpt07.

\section{AMO Lorentz tests of the Minkowski limit of the SME}

\subsection{Couplings of fermions to the SME background}
Since Lorentz-violating background fields
are known to be small,
the analysis of effects that might occur is readily handled
using perturbation theory.
For most applications with ordinary fermionic matter,
the unperturbed system is obtained from the Dirac equation
with solutions being the spinors $\ps$.
Many of the principles encapsulated in the Dirac equation
have recently been studied
with an eye towards atomic-interferometry based tests
of basic principles including Lorentz symmetry,
the universality of free fall,
locality,
and the superposition principle
\cite{cl}.
Other issues such as stability and causality
have been researched in this context,
as well as in field theory \cite{causality}.
A variety of couplings of fermions to Lorentz-violating background fields
have been studied.
For example, one term appearing in the lagrangian density is
\cite{dcak97}:
\beq
\cL_{\mbox{\footnotesize matter}} \supset b_\mu\ps \ga_5\ga^\mu \ps
\, .
\eeq
Distinct coefficients $b_\mu$ are used to quantify
Lorentz violation for each fermion.
Perturbative analysis of this term
can be used to find the shifts in the spectra of
electrons in Penning traps \cite{penntrap},
hydrogen and antihydrogen \cite{bkrHbar},
atomic clocks and masers\cite{clockmaser},
torsion pendula \cite{tpend},
and other systems.
Many of these experiments involve
the comparison of highly stable frequencies
with each other
\cite{cc}.

In the neutron sector of the SME,
the most stringent bounds on Lorentz violation
have been obtained using a He-Xe dual maser
at the Harvard-Smithsonian Center for Astrophysics.
Limits on symmetry breaking among
the rotational components of the Lorentz group
are at the level of
$10^{-31}$ GeV \cite{Bear00}
and, on the boost components,
at the level of
$10^{-27}$ GeV \cite{Cane04}.
An improvement in precision of about an order of magnitude
is expected after current upgrades are completed.
These include
upgraded temperature controls,
optimized noble gas pressures and cell geometries,
increased Zeeman frequency,
proper spatial definition of masing ensembles,
and improved stability of the double-tuned resonator
\cite{HeXe}.

A group at Princeton University
has designed, built, and operated a potassium-helium co-magnetometer
with sensitivity to electron, proton, and neutron coefficients for Lorentz violation.
The potassium and helium atoms are confined within a glass cell
and controlled using optical pumping techniques.
Using data taken over a period of 15 months,
this magnetometer, dubbed CPT-I,
has achieved excellent sensitivity to a variety of effects
including sidereal signals that would be expected from a fixed Lorentz-violating background.
Preliminary results include a bound at the level of about
$10^{-30}$ GeV on the equatorial components of the proton $b_\mu$
coefficient \cite{mr}.
A second-generation co-magnetometer, CPT-II,
is currently being implemented to achieve yet higher sensitivities.
This device is mounted on a turntable,
making possible cycle times of much less than a day.
This is expected to much improve the sensitivity to sidereal effects.
Other innovations have been introduced to
improve sensitivities in various ways.
These include shorter optical path lengths,
reduction of convection noise in the oven area,
evacuation of air from the optical path,
and improved magnetic shielding.
CPT-II is expected to
surpass the sensitivity of CPT-I by several orders of magnitude
\cite{ss}.

Experiments with antihydrogen
have the potential to find signals of Lorentz violation
that are not accessible with other systems.
There are three groups working on antihydrogen physics at CERN:
the `Antihydrogen Laser Physics Apparatus' (ALPHA) collaboration,
the `Atomic Spectroscopy and Collisions using Slow Antiprotons' (ASACUSA) collaboration ,
and the `Antihydrogen Trap' (ATRAP) collaboration.
ALPHA and ASACUSA were represented at \cpt07.

The ASACUSA collaboration has conducted several precision experiments
using the laser spectroscopy of antiprotonic helium.
The group has measured the antiproton-to-electron mass ratio
to a precision of 2 parts per billion \cite{asacusa06},
which is within an order of magnitude
of the proton-to-electron mass ratio found using a Penning-trap
comparison of a proton and an electron.
Theoretical studies of Lorentz violation in antihydrogen
have shown that unsuppressed signals could potentially
occur in the comparison of the hyperfine spectral lines of hydrogen
and antihydrogen \cite{bkrHbar}.
The ASACUSA collaboration plans
to measure the hyperfine lines of
antihydrogen in a Stern-Gerlach beam arrangement \cite{juhasz}.
The expected resolution is at the level of $10^{-21}$ GeV.

The ALPHA collaboration aims to produce trapped antihydrogen
with the eventual goal of conducting precise comparisons
of the spectra of antihydrogen and hydrogen.
The group demonstrated the trapping of antiprotons from the CERN
antiproton decelerator in 2006.
The design involves a Penning-Malmberg trap
featuring a magnetic octopole configuration \cite{alpha07}
to confine positrons and antiprotons in the same region.
Methods of cooling and compressing the plasmas
to enhance the rate of
antihydrogen formation are being investigated.

The E\"ot-Wash group at the University of Washington in Seattle
has investigated couplings of spin to Lorentz-violating SME
background fields \cite{heckel06}.
The apparatus used for this
consists of a spin-polarized torsion pendulum
suspended by a 75-cm tungsten fiber.
It has minimal magnetic and gravitational moments,
and a net number of polarized spins on the order of
$10^{23}$,
making it highly sensitive to
the coupling of these spins to
the Lorentz-violating background field $b_\mu$ for the electron.
The component of this background
that is parallel to the rotation axis of the Earth
has been bounded at the level of a few parts in $10^{-30}$ GeV
by this experiment.
The limits it places on the two components in the equatorial plane
are an additional order of magnitude tighter.

Another system where large numbers of spin-polarized atoms
may be able to amplify Lorentz-violating effects
is the Bose-Einstein condensate.
Since this involves atoms that are bosonic,
the statistical properties
can be expected to be very different from fermionic systems.
Under suitable conditions,
spin-polarized Bose-Einstein condensates
may be sensitive to Lorentz-violating background fields
at a level comparable to other existing tests
\cite{dcpm}.

Space-based experimental tests of fundamental physics
are motivated by their potential to reach higher precisions than earth-based ones
and to probe otherwise inaccessible observables.
A number of proposals and projects at various stages of development
exist in the European Space Agency and NASA communities.
These include the Laser Interferometer Space Antenna (LISA),
its precursor LISA Pathfinder (LISAPF),
the Grand Unification and Gravity Explorer (GAUGE),
the Laser Astrometric Test of Relativity (LATOR),
the Astronomical Space Test of Relativity using Optical Devices (ASTROD),
the Odyssey Mission aimed at exploring gravity in the Solar System,
and the Matter-Wave Explorer of Gravity (MWXG) \cite{ts}.
Technological advances making such missions attractive for physics experiments
include the ability to create drag-free platforms using systems such as micronewton
thrusters,
and precision capacitive, magnetic, and optical sensing of proof-mass behavior.
Other proposals for high-precision space tests include ones based on
atomic-clock comparisons \cite{bklr2003}.

A recent muon experiment at the Brookhaven National Laboratory,
while not directly in the field of AMO physics,
may be of interest since it has many similarities with
the Penning-trap system.
The E821 experiment, run by the $g-2$ collaboration,
measured the anomaly frequency of positive and negative muons stored in the
AGS ring.
Analysis of sidereal variations in these frequencies
limited the equatorial-plane $\bt_\mu$ coefficients for Lorentz violation
at the level of $10^{-24}$ GeV
\cite{bennett07}.
Further analysis is expected to be able to place constraints on
a variety of combinations of muon coefficients
for Lorentz violation.

\subsection{Couplings of photons to the SME background}
In the Minkowski limit of the SME without torsion,
the following photon-sector term
has been the primary focus of a number of experiments:
\beq
\cL_{\mbox{\footnotesize matter}} \supset
- \frac 1 4 (k_{F})_{\ka\la\mu\nu}F^{\ka\la} F^{\mu\nu}
\, .
\eeq
To date, most of the tests in this sector
have focussed on propagating electromagnetic fields,
although practical tests are possible in statics
\cite{qbakstatics}.
After accounting for symmetries
there are nineteen independent components for the coefficients
$(k_{F})_{\ka\la\mu\nu}.$
There are ten linear combinations of these
that imply birefringence,
and these have been constrained tightly using observations
of distant cosmological sources \cite{akmm01}.
The remaining nine have been studied
extensively in laboratory experiments
with microwave and optical cavity oscillators.
Experiments have placed bounds on
linear combinations of these nine coefficients,
denoted by $\kt_{e-}$ and $\kt_{o+}$.
Recent results from two such experiments
were presented at \cpt07.

An experiment at the University of Western Australia
involves two cryogenic sapphire
oscillators, rotated about the vertical axis
with a period of 18 seconds.
By taking data over a time scale of about one year,
measurements have been made
of all eight independent $\kt_{e-}$ and $\kt_{o+}$
components without any non-cancelation assumptions
\cite{mt}.
A second experiment by this group at the University of Western Australia
consists of a Mach-Zehnder microwave interferometer mounted on a rotating platform.
After completion of the development stages, it is expected to
measure $\kt_{\mbox{\footnotesize tr}}$ at competitive
levels \cite{mt}.

An order of magnitude improvement in sensitivity
is expected in an experiment
at the Humboldt University in Berlin, Germany.
It compares the optical frequencies in two orthogonal
cavities created in
a single block of fused silica \cite{ap}.
The system is maintained in a thermally insulated
and vibration isolated vacuum chamber, which
is mounted on a turntable with a period of 45 seconds.
This and other experiments have utilized such rotating turntables
to improve precisions over earlier versions that
relied on the rotation of the earth to seek anisotropies.
The preliminary results of this experiment
place some of the tightest constraints on a variety
of the $\kt_{e+}^{JK}$ and $\kt_{o-}^{JK}$
coefficients for Lorentz violation \cite{LVtables}
and are listed in Table~\ref{kTable}.
\begin{table}
\caption{Photon-sector results reported by
the Humboldt University group \cite{ap}. The value of $\be$ is $10^{-4}$}								
\begin{tabular}{@{}cr@{\hspace{12pt}}@{}}
& \\
\hline
&\\	
{\bf Combination}	&	\multicolumn{1}{c}{{\bf Result}} \\[2 pt]
$	(\kt_{e-})^{XY}	$&$	(-0.1 \pm 0.6) \times 10^{-17}	$ \\ [1pt]
$	(\kt_{e-})^{XZ}	$&$	(-2.0 \pm 0.9) \times 10^{-17}	$ \\ [1pt]
$	(\kt_{e-})^{YZ}	$&$	(-0.3 \pm 1.4) \times 10^{-17}	$ \\ [1pt]
$	(\kt_{e-})^{XX} - (\kt_{e-})^{YY}	$&$	(-2.0 \pm 1.7) \times 10^{-17}	$ \\ [1pt]
$	(\kt_{e-})^{ZZ}	$&$	(-0.2 \pm 3.1) \times 10^{-17}	$ \\ [1pt]
$	\be (\kt_{o+})^{XY}	$&$	(-2.5 \pm 2.5) \times 10^{-17}	$ \\
$	\be (\kt_{o+})^{XZ}	$&$	(1.5 \pm 1.7) \times 10^{-17}	$ \\
$	\be (\kt_{o+})^{YZ}	$&$	(-1.0 \pm 1.5) \times 10^{-17}	$ \\
&\\	
\hline
\label{kTable}
\end{tabular}								
\end{table}								

The results from geographically distant experiments,
such as the ones discussed above,
can be combined to obtain additional information.
Furthermore,
coordinate and field redefinitions
can be used to
establish various links between results in different sectors of the SME.
Recent work along these lines has led to several results
\cite{hm}.

Theoretical considerations of precision Doppler-shift experiments
show that sensitivity to some coefficients
for Lorentz violation,
such as the $c_{\mu\nu}$ for protons and electrons,
is possible in principle
\cite{lane05}.
An experiment at the Max Planck Institute for Nuclear Physics in Heidelberg,
Germany,
has conducted a test of special relativity
by measuring the Doppler shifts of beams of lithium atoms
traveling at speeds of about 3 to 6 \% of the speed of light
\cite{Reinhardt07}.
One of the results of this experiment is
\beq
|\kt_{\mbox \footnotesize tr}| < 8.4 \pm 10^{-8}
\, ,
\eeq
a bound on Lorentz violation in the photon sector.
Additional sensitivity,
to components in the fermion sector,
may be possible through the use of circularly polarized lasers.

Higher-order couplings constructed from the electromagnetic fields
$A_\mu$ or $F^{\mu\nu},$
and derivatives,
have been studied recently \cite{akmm07}.
They include,
for example,
the term
\beq
\cL_{\mbox{\footnotesize matter}} \supset - \frac 1 2 \ep^{\ka\la\mu\nu}
(k_{AF}^{(5)})_\ka^{\phantom{\ka}\ga\ta} A_\la
\partial_\ga \partial_\ta F_{\mu\nu}
\, .
\eeq
The constant coefficient $(k_{AF}^{(5)})_\ka^{\phantom{\ka}\ga\ta} $
has the dimension of inverse mass,
which ensures that the full term is of dimension four in the mass.
In general,
there is an infinite number of terms constructed in this way
by the inclusion of further derivatives.
Effects of such terms include vacuum birefringence,
and recent work has placed limits on the coefficients
$(k_{AF}^{(5)})_\ka^{\phantom{\ka}\ga\ta}$
and higher-order coefficients
by studying polarization data from observations of
the cosmic microwave background \cite{akmm07}.

\section{AMO Lorentz tests in the gravitational sector}
\subsection{Pure gravity sector}
The first constraints on pure-gravity sector SME coefficients
were presented at \cpt07 by two experimental groups,
one working with lunar-laser ranging,
and the other with atomic interferometry.
The SME terms of interest in these experiments
appear in the lagrangian density in the form \cite{qbak06}
\beq
\cL_{\mbox{\footnotesize gravity}} \supset
\frac 1 {16 \pi G} s^{\mu\nu} R^T_{\mu\nu}
\, ,
\eeq
where $G$ is the universal gravitational constant.
This term couples the traceless Ricci tensor $R^T_{\mu\nu}$,
obtained by contraction of
the curvature tensor $R_{\mu\nu\al\be}$,
to a Lorentz-violating background
expressed as $s^{\mu\nu}$.
The coefficients $s^{\mu\nu}(x)$
have vacuum expectation values $\sb^{\mu\nu}$
induced by spontaneous violation of local Lorentz symmetry.
The fluctuations of fields like $s^{\mu\nu}(x)$
about the vacuum expectation values $\sb^{\mu\nu}$
have fascinating implications
for physics
\cite{modes}.
The coefficients of interest at present are
$\sb^{\mu\nu}$, which are traceless and antisymmetric
and so have 9 independent values.

A group at Stanford University
has used a highly sensitive atomic gravimeter
to place bounds on
combinations involving the $\sb$ coefficients
and photon-sector coefficients.
The outstanding precision of gravimeters based on atom interferometry
stems from
the ability of neutral atoms to approach a freely falling reference frame
with high accuracy,
and the ability of lasers to interrogate the motion
with fantastic precision.
The Stanford group controls and measures the
behavior of matter waves formed using clouds of Cs atoms.
The device has resolved
the acceleration of gravity
more than three times better
than the best previously reported value
\cite{muller07}.
Their results are given in Table~\ref{gravimeterTable}.

\begin{table}
\caption{Pure-gravity sector results reported by the Stanford University group,
using a Cesium-based atomic gravimeter \cite{muller07}. The $\si$ coefficients
are a combination of pure-gravity sector $\sb$ coefficients
and photon-sector coefficients.}								
\begin{tabular}{@{}cr@{\hspace{12pt}}@{}}
& \\
\hline
&\\	
{\bf Combination}	&	\multicolumn{1}{c}{{\bf Result}} \\[2 pt]
$	\si^{XX}-\si^{YY}	$&$	(-5.6 \pm 2.1)\times 10^{-9}	$\\	[2pt]			
$	\si^{XY}	$&$	(-0.09 \pm 79) \times 10^{-9}	$\\	[2pt]			
$	\si^{XZ}	$&$	(-13 \pm 37) \times 10^{-9}	$\\	[2pt]			
$	\si^{YZ}	$&$	(-61 \pm 38) \times 10^{-9}	$\\	[2pt]			
$	\si^{TY}	$&$	(-2.0 \pm 4.4) \times 10^{-5}$\\	[2pt]			
$	\si^{TX}	$&$	(5.4 \pm 4.5) \times 10^{-5}$\\	[2pt]			
$	\si^{TZ}	$&$	(1.1 \pm 26) \times 10^{-5}$\\				
&\\	
\hline
\label{gravimeterTable}
\end{tabular}								
\end{table}								

A Harvard group
presented results of an analysis of more than 30 years of
lunar laser-ranging data,
constraining six independent combinations of $\sb$ coefficients
at the level of $10^{-6}$ and $10^{-11}$.
These data were collected primarily at the
McDonald Laser-Ranging Station in Texas, USA,
and
the C\^ote d'Azur station in Grasse, France,
during the period spanning September 1969
and December 2003.
Their results are reported in Ref.\ \cite{battat},
and are summarized in Table~\ref{apolloTable}.
The Apache Point Observatory Lunar Laser-ranging Operation (APOLLO)
\cite{apollo}
is expected to improve on these results by about an order of magnitude.
The telescope at Apache Point in New Mexico, USA,
can detect reflections from the lunar retroreflector arrays
even in daylight conditions,
and ranging can be achieved at the millimeter level.

\begin{table}
\caption{Pure-gravity sector results reported by the Harvard-Smithsonian group,
based on archival lunar laser-ranging data \cite{battat}.}
\begin{tabular}{@{}cr@{\hspace{12pt}}@{}}
& \\
\hline
&\\	
{\bf Combination}	&	\multicolumn{1}{c}{{\bf Result}} \\[2 pt]
$	\sb^{11}-\sb^{22}	$&$	(1.3 \pm 0.9) \times 10^{-10}	$\\	[2pt]			
$	\sb^{12}	$&$	(6.9\pm 4.5) \times 10^{-11}	$\\	[2pt]			
$	\sb^{02}	$&$	(-5.2 \pm4.8) \times 10^{-7}	$\\	[2pt]			
$	\sb^{01}	$&$	(-0.8\pm 1.1) \times 10^{-6}	$\\	[2pt]			
$	\sb_{\Om_\oplus c}	$&$	(0.2 \pm 3.9) \times 10^{-7}	$\\	[2pt]			
$	\sb_{\Om_\oplus s}	$&$	(-1.3\pm 4.1) \times 10^{-7}	$\\				
&\\	
\hline
\label{apolloTable}
\end{tabular}								
\end{table}								

\subsection{Couplings of matter with gravitational fields}
Terms in
$\cL_{\mbox{\footnotesize matter}}$
that couple matter to gravitational fields
are currently being studied \cite{akjt}
since they offer the possibility of obtaining new sensitivities
to Lorentz violation in,
for example,
the fermion sector.
One approach is to start from the relativistic theory
using the spin connection and the vierbein \cite{akgrav04}
as the basic gravitational objects,
and the Dirac fermion $\ps$ and the photon field $A^\mu$
as the basic non-gravitational objects.
One can then extract the nonrelativistic limit
using,
for example,
a Fouldy-Wouthuysen transformation,
to obtain a formalism appropriate for direct experimental analysis.
Another approach of interest is the classical theory
involving point-particles
rather than wave functions.
Results are expected to provide the first direct sensitivities
to the $a_\mu$ coefficients for the proton, neutron, and electron
\cite{akjt}.

Torsion is a basic field in Riemann-Cartan theories of gravity,
giving twisting degrees of freedom that are distinct from
curvature.
This field $T^\mu_{\phantom{\mu}\al\be}$, which has 24 independent components,
can be nonzero even in the Minkowski flat-spacetime limit.
Couplings of fermions and other particles to this field
have similarities with couplings of fermions
to Lorentz-violating background fields in the SME.
This fact has been exploited recently to
deduce new bounds on 15 of the torsion components
and the most stringent bounds on the four
minimally-coupled torsion components
\cite{aknrjt}.
The latter four bounds,
on the axial components $A^\mu \equiv \ep^{\al\be\ga\mu} T_{\al\be\ga}/6$,
are:
\bea
|A_T| < 2.9 \times 10^{-27} \mbox{GeV} \, , &\phantom{xxxxx}&
|A_X| < 2.1 \times 10^{-31} \mbox{GeV} \, , \nonumber\\
|A_Y| < 2.5 \times 10^{-31} \mbox{GeV} \, , &&
|A_Z| < 1.0 \times 10^{-29} \mbox{GeV} \, . \nonumber
\eea
These results are based on experiments
with a spin-polarized torsion pendulum \cite{heckel06},
and with a helium-xenon dual maser \cite{Cane04}.

\section{Closing}
The Standard-Model Extension is an umbrella framework for
tests of Lorentz symmetry in nature.
By setting up a general coefficient space
for all Lorentz violations
it
has allowed new tests of Lorentz symmetry to be
identified across the sectors of physics,
and
made possible the comparison of Lorentz tests
from vastly differing systems.
Experiments,
many of them in the sphere of AMO physics,
have delved into
the SME coefficient space for the last decade.
On the theoretical front,
recent research has focused on the gravitational sector
of the SME.
This article reports primarily on AMO Lorentz-symmetry tests
featured in presentations made at
the \cpt07 meeting
held in Indiana in August 2007 \cite{cpt07}.
Included are bounds on
a number of coefficients measured for the first time in the pure gravity sector.
This sector of the SME is likely to
generate further experimental activity
as a number of unexplored regions offer
the alluring prospect of finding Lorentz violations.

\end{document}